\documentclass[11pt,twoside,letterpaper]{article} 
\usepackage{times,fancyhdr}
\usepackage[dvips]{graphicx}
\makeatletter 
\def\cleardoublepage{\clearpage\if@twoside \ifodd\c@page\else%
    \hbox{}%
    \thispagestyle{empty}%
    \newpage%
    \if@twocolumn\hbox{}\newpage\fi\fi\fi} 
\makeatother 
\renewcommand{\thesection}{\arabic{section}.}
\renewcommand{\thesubsection}{\thesection\arabic{subsection}.}

\newcommand{\beq}{\begin{equation}}
\newcommand{\eeq}{\end{equation}}
\newcommand{\beqn}{\begin{eqnarray}}
\newcommand{\eeqn}{\end{eqnarray}}
\newcommand{\btab}{\begin{tabular}}
\newcommand{\etab}{\end{tabular}}
\newcommand{\etal}{{\em{et al.}}}

\def\figurename{Figure}
\makeatletter
\renewcommand{\fnum@figure}[1]{\figurename~\thefigure.}
\makeatother
\def\tablename{Table}
\makeatletter
\renewcommand{\fnum@table}[1]{\tablename~\thetable.}
\makeatother
\begin{document}
\title{
{\begin{flushleft}
\vskip 0.45in
{\normalsize\bfseries\textit{Chapter~1}}
\end{flushleft}
\vskip 0.45in
%
%
%
%
\bfseries\scshape Equation of State for Proto-Neutron Star}}
\author{\bfseries\itshape G. Shen\thanks{E-mail gshen@lanl.gov}\\
Theoretical Division, Los Alamos National Laboratory, Los Alamos, NM, 87545, USA}
\date{}
\maketitle
\thispagestyle{empty}
\setcounter{page}{1}
\thispagestyle{fancy}
\fancyhead{}
\fancyhead[L]{In: Neutron Star Crust \\ 
Editors: C.A. Bertulani and J. Piekarewicz, pp. {\thepage-\pageref{lastpage-01}}} 
\fancyhead[R]{ISBN 0000000000  \\
\copyright~2012 Nova Science Publishers, Inc.}
\fancyfoot{}
\renewcommand{\headrulewidth}{0pt}
\vspace{2in}
\begin{abstract}

Physics of the Equation of State (EoS) for proto-neutron star (PNS) concerns properties of neutron rich matter at finite temperature over wide range of densities. In this contribution we discuss the structure of PNS inner crust in a relativisitc mean filed model with spherical Wigner-Setiz approximation, and the composition of matter around neutrino-sphere in PNS in a virial expansion of non-ideal gas composed of nucleons and nuclei. We go on to discuss several new complete EoS for PNS and supernova, whose detailed composition is important for the neutrino dynamics. We focus on one important distinction for various EoS - the density dependence of symmetry energy $E'_{sym}$, and its interesting correlation with the radii of neutron star, as well as properties of neutron distribution in neutron rich nuclei. Improved understanding of $E'_{sym}$ from terrestrial experiment on neutron distribution of neutron rich nuclei, benchmark calculations via ab initio methods, and statistical analysis on good quality observational data will advance our knowledge on EoS.

\end{abstract}

\noindent \textbf{PACS} 21.65.Mn,26.50.+x,26.60.Kp,21.60.Jz
\vspace{.08in} \noindent \textbf{Keywords:} Proto-Neutron Star; Equation of State of Nuclear Matter.
%
\pagestyle{fancy}
\fancyhead{}
\fancyhead[EC]{G. Shen}
\fancyhead[EL,OR]{\thepage}
\fancyhead[OC]{Equation of State for Proto-Neutron Star}
\fancyfoot{}
\renewcommand\headrulewidth{0.5pt} 
%

\section{Introduction}

Proto-neutron star (PNS) is born subsequent to gravitational core collpase supernova, which explodes once every second in our universe. In this chapter we describe the equation of state (EoS) of nuclear matter for PNS, particularly the strcuture of inner crust and composition around neutrino-sphere, which is the last scattering site before neutrinos leave PNS. The EoS is tied to many important questions in nuclear physics and astrophysics: what is the biggest and largest neutron star that nature could make? what's the role of neutrino-matter interaction in supernova explosion dynamics? what's the spectrum of neutrinos emerged from neutrino-sphere and eventually observable in future terrestrial detectors? where did the chemical elements beyond iron come from?

The EoS for hot, dense matter relates energy and pressure to temperature, density, and composition. The properties of hot dense matter, for example its pressure at high baryon density - larger than normal nuclear density 3$\times 10^{14}$ g/cm$^3$, have been the focus of many extradinary terrestrial experiments, including heavy ion collisions with Au \cite{heavyions}. The pressure of nuclear matter at high density determines how large a neutron star our nature could realize.

The properties of nuclear matter depends on its composition, particularly the proton-neutron number asymmetry, or iso-spin dependence, conveniently characterized by the parameter called (a)symmetry energy $E_{sym}$. Most stable nuclei have a small such asymmetry and tell us little about how the EoS changes with the asymmetry. The neutron/proton rich isotopes will be studied with new tools like the Facility for Rare Isotope Beams (FRIB) \cite{frib}, a heavy ion accerelator to be built in Michigan State University. Studies by these new tools will help us understand when the nuclei would become unstable upon too many neutrons (or protons) added, and ultimately tell us the composition of nuclear matter in PNS given temperature, density, and proto-neutron asymmetry.

Density dependence of symmetry energy $E'_{sym}$ is one key unknown in nuclear physics and nuclear astrophysics, where neutron rich matter is particularly relevant.  There are many interesting correlations with  $E'_{sym}$ that have been studied in recent years. The pressure of nuclear matter is proportional to $E'_{sym}$. With a higher pressure if  $E'_{sym}$ is large , neutron rich nucleus, such as $^{208}$Pb, is found to have a larger neutron radius \cite{Brown00}. This has motivated the Lead Radius Experiment (PREX) \cite{prex} to accurately measure the neutron radius in $^{208}$Pb with parity violating electron scattering \cite{parity}. On the other hand, the radius of a canonical 1.4 solar mass neutron star is determined by the nuclear matter at similar density inside $^{208}$Pb, therefore a larger pressure at such density tends to give a bigger radius for 1.4 solar mass neutron star \cite{HP01}. Current large uncertainties in the EoS lead us to generate several big tables of EoS based on dinstinct properties of nuclear matter at high densities, which could be used in astrophysical simulations such as PNS evolution to identify astrophysical observables with related nuclear matter properties.

The dynamical response of nuclear matter, important for the evolution of PNS, depends on its detailed composition. In the inner crust of PNS, nuclei with unconventional shapes such as plate or rod like could appear \cite{pasta1}. Neutrinos could coherently scattering off these novel nuclei - nuclear pasta, because such nuclear shapes have sizes comparable to neutrino de-Broglie wavelength in PNS \cite{pasta2}. Moreover, light nuclei could be abundant around neutrino-sphere, therefore influence the spectra of neutrinos emerged from neutrino-sphere \cite{light07}. Finally, the neutrino spectra from PNS determine the neutron-proton ratio in the neutrino driven wind from supernova, which is one most promising site for neutron rich nuclei to be synthesized \cite{rp}.

The chapter is organized as follows. In section 2 we describe the structure of PNS inner crust in a relativistic mean field model with spherical Wigner-Seitz apprximation for the lattice. In section 3 we describe the EoS around neutrino sphere in a virial expansion of nonideal gas of nucleons and nuclei. In section 4 we discuss several new complete EoS tables covering large range of densities, temperatures, and proton fractions, to be used in astrophysical simulations. In the following section 5, we discuss one important distinction among the various EoS tables, the density dependence of symmetry energy at high density and its correlation with various observables. Fianlly we conclude in section 6.

\section{Structure of Inner Crust}

Realistic calculations with modern nuclear interactions for the structure of neutron star crust started with the seminal work of Negele and Vautherin, who applied a Skyrme Hartree-Fock calculation with two-body potentials \cite{Negele}, and obtained the ground state within the spherical Wigner-Seitz (WS) approximation.  In WS approximation, the
unit cell of a crystal lattice is modeled as a sphere.  They found, as the system became more
neutron rich and the density increased, neutrons escaped and the system approached a uniform state near nuclear density.  Since then there have been many investigations \cite{more} with more sophisticated interactions and more complicated lattice configurations, such as rods or plates.  These non-spherical `pasta' phases, which seem to appear within a significant range of sub-nuclear densities are relevant for the structure of neutron star crusts \cite{crust1,crust2} and the dynamics of supernovae \cite{sn}. For simplicity, this contribution is limited to discuss perfect crystal with a single nuclear species at lattice sites (for the possibility of heteronuclear compounds, see Ref. \cite{Hetero} and references therein).

In proto-neutron star, neutrinos are trapped for tens of seconds in the hot, dense nuclear medium. The proton fraction in thermal beta equilibrium matter therefore evolves as neutrinos diffuse out of proto-neutron star. As a result, one has to consider a large range of proton fractions, as well as temperatures, in the equation of state for PNS.  For hot nuclear matter, assuming that the spherical WS cell
still remains a good approximation to the realistic lattice structure,
people have studied the EoS with different models. Using a phenomenological compressible
liquid-drop model, Lattimer and Swesty \cite{LS} (L-S) produced an
equation of state for hot dense matter that has been widely used in supernova simulations.  Later, H. Shen \etal \cite{Shen98a} (S-S) constructed an equation of state based on
Thomas-Fermi and variational approximations to a relativistic mean field (RMF) energy functional.  However neither method takes into account the shell structure of finite nuclei or explores the full range of density distributions possible even in the spherical WS approximation.

In recent years, RMF models have
provided a consistent description for the ground state properties
of finite nuclei, both along and far away from the valley of beta
stability \cite{HS1,HS2,SW,Rein,Ring}.  These models incorporate
the spin-orbit splitting naturally and the relativistic formalism
provides a framework to extrapolate  the properties of non-uniform
and uniform nuclear matter to high densities.  Furthermore there
is a close relation between phenomenological RMF models, that
simply fit parameters to properties of finite nuclei, and more
systematic effective field theory approaches that enumerate all of
the possible interactions allowed by symmetries.

In this section, we use an RMF model for non-uniform matter at
intermediate density (and can be easily extended to uniform matter at high density).  Low density pure neutron matter is analogous to a unitary gas \cite{HS05b}, where the neutron-neutron scattering length is
much larger than both the effective range of nuclear force and the average inter-particle spacing. To better describe neutron-rich matter at low density, we use a density dependent scalar
meson-nucleon coupling. At high density, the model reduces to the normal RMF parameter set NL3 \cite{NL3}.  The unit lattice of non-uniform nuclear matter is conveniently
approximated by a spherical WS cell. The meson mean
fields and nucleon Dirac wave functions inside the Wigner-Seitz cell are solved
fully self-consistently. The
size of the WS cell is found by minimization of the free energy per
nucleon.  The WS approximation provides a framework to incorporate the best known microscopic nuclear physics \cite{Negele}. The nuclear shell structure effects
are included automatically and it is already possible for
some effects of complex nuclear pasta states to be included in spherical calculations in the form of shell states \cite{HS08}.  Full three-dimensional WS calculations, by various nuclear models including liquid droplet models, quantum molecular dynamic simulations, and Hartree-Fock calculations, predict a whole sequence of pasta phases \cite{Gogelein,pasta1,pasta2}, which
would make the transition to uniform matter more smooth. However
this will demand increased complexity and much larger computational resources if one wishes to generate complete EoS for PNS. For simplicity in this contribution we use the spherical WS approximation.

\subsection{Non-uniform nuclear matter in Wigner-Seitz approximation}
\label{subsec.nonuniform}

The formalism for relativistic mean field theory has been reviewed
in previous works, see eg \cite{SW,Rein,Ring,SHT10a,HS08}. The basic ansatz of the RMF theory is a Lagrangian density where
nucleons interact via the exchange of sigma- ($\sigma$), omega-
($\omega_\mu$), and rho- ($\rho_\mu$) mesons, and also photons
($A_\mu$). \beqn\label{lagrangian}
    {\cal L}&=&\overline{\psi} [i{\gamma^\mu}
              {\partial_\mu}-m-{\Gamma_\sigma}\sigma - g_\omega
              \gamma^\mu\omega_\mu \nonumber\\
              &&-\
              g_\rho \gamma^\mu \vec{{\bf
              \tau}}\cdot \vec{\mbox{\boldmath$\rho$}}_\mu - e\gamma^\mu\frac
              {1+\tau_3}{2} A_\mu ]\psi\nonumber\\
            &&+\
            \frac{1}{2}\partial^\mu\sigma\partial_\mu\sigma-\frac{1}{2}m_\sigma^2\sigma^2-
              \frac{1}{3}g_2\sigma^3\ -\ \frac{1}{4}g_3\sigma^4\nonumber\\
            &&-\
              \frac{1}{4}\omega^{\mu\nu}\omega_{\mu\nu}+\frac{1}{2}m_\omega^2\omega^\mu\omega_\mu+
              \frac{1}{4}c_3\left(\omega^\mu\omega_\mu\right)^2\nonumber\\
            &&-\ \frac{1}{4}\vec{\mbox{\boldmath$\rho$}}^{\mu\nu}
              \cdot\vec{\mbox{\boldmath$\rho$}}_{\mu\nu}+
              \frac{1}{2}m_\rho^2\vec{\mbox{\boldmath$\rho$}}^\mu
              \cdot\vec{\mbox{\boldmath$\rho$}}_\mu -\
              \frac{1}{4}A^{\mu\nu}A_{\mu\nu}
\eeqn 
Here the field tensors of the vector mesons and the
electromagnetic field take the usual forms. To fit
neutron rich matter at low density from ab initio methods, we introduce a density dependent coupling
between the scalar meson and the nucleon, $\Gamma_\sigma = \Gamma_\sigma(n)$
($n\equiv\sqrt{j_\mu j^\mu}$ is baryon density and $j_\mu$ is nucleonic current). 

In charge neutral nuclear matter composed of neutrons, $n$,
protons, $p$, and electrons, $e$, there are equal numbers of
electrons and protons.  Electrons can be treated as a uniform
Fermi gas at high densities {\footnote{It needs electron density
$> 10^6$ g/cm$^3$, which is easily surmounted in the regime considered in the inner crust.}}. They contribute to the Coulomb energy of
the $npe$ matter and serve as one source of the Coulomb potential in nuclear mean field.

The variational principle leads to the following equations of
motion 
\beqn\label{nucleon-motion}
     [\mathbf{\sl{\alpha}}\cdot\mathbf{\sl{p}}\ +\ V(\mathbf{r})\ +\
     \sl{\beta} (m\ +\ S(\mathbf{r}))] \psi_i\ =\ \varepsilon_i\psi_i
\eeqn for the nucleon spinors, with vector and scalar potentials
\beqn \label{Dirac}
     \begin{array}
        {l} V(\mathbf{r}) =\ \beta \{g_{\omega}\rlap{/}{\omega}_{\mu}
        + {g_{\rho}} \vec{\tau}\cdot
        \vec{\rlap{/}\rho_{\mu}} + e \frac{ (1 + \tau_3)}{2}\ \rlap{/}A_{\mu} + \Sigma^R \}, \\
        S(\mathbf{r}) =\ \Gamma_{\sigma}\sigma, \
     \end{array}
\eeqn 
where \beq \Sigma^R = \frac {\gamma^\mu j_\mu} {n}
\frac{\partial\Gamma_\sigma}{\partial n} \rho_s \sigma, \eeq is
the rearrangement term due to the density dependent coupling between the 
sigma meson and the nucleon, and $\rho_s$ is the scalar density of
nucleons. The form of $\Sigma^R $ is adjusted to reproduce 
ab initio neutron matter calculation at low density \cite{SHT10b}.

In the WS approximation, the lattice Coulomb energy consists of
contributions from neighboring unit cells. This correction to the WS
Coulomb energy is important for determining the stable
configuration of WS cells and the transition density to uniform
matter, especially when the system has a large proton fraction.
We include the exact Coulomb energy in calculating the
free energy of WS cell of radius $R_c$.  Following the treatment in
Ref.~\cite{Kittel,Oyamatsu84}, we calculate the Coulomb energy per
unit cell as, \beq\label{ExactCoulomb} W_c = \frac 1 2 \sum_{\vec{G}} ' \frac{I_{hkl}^2}{a^3
\vec{G}^2}, \eeq where $a$ is lattice constant defined by
$a^3=V_{cell}=\frac{4\pi} 3 R_c^3$, and \beq \vec{G} =
h\vec{A}+k\vec{B}+l\vec{C}. \eeq Here $h, k$, and $l$ are
integers, and $\vec{A}$, $\vec{B}$, and $\vec{C}$ are the
primitive transformation vectors of the reciprocal lattice. The
prime on the sum means that the point $\vec{G}$ = 0 is excluded.
The form factor $I_{hkl}$ is given by \beq I_{hkl} = \int_{cell}
\rho_p(\vec{r}) e^{-i\vec{G}\cdot\vec{r}} d\vec{r}. \eeq Oyamatsu
\etal \cite{Oyamatsu84} assumed a uniform proton density inside the nucleus and found that the stable configuration is a Body-Centered Cubic (BCC) lattice.  Using realistic proton density distribution, we also find that a BCC lattice gives the lowest Coulomb energy.
The electron charge screening effects are negligible since the electron Thomas-Fermi screening length is larger than the lattice spacing in the inner crust \cite{screening}. There are also small corrections due to electron-exchange interactions and electron polarizations, which are discussed in Ref. \cite{coulomb0}.

In the spherical Wigner-Seitz approximation, the unit cell of a crystal lattice, complex close-packed polyhedron, is approximated by a
spherical cell of the same volume. One WS cell has one nucleus.  We apply boundary conditions on the wave functions at the edge of WS unit cell.  To achieve a uniform density distribution for a free neutron gas, we require
that at the cell radius, all wave functions of even parity vanish,
and the radial derivative of odd parity wave functions also
vanishes \cite{Negele}.

\subsection{Solutions for Wigner-Seitz cells}


For nonuniform matter we find new shell states which minimize the free
energy per baryon over a significant density range. Shell states
have inside and outside surfaces and they can minimize the Coulomb
energy of high $Z$ (large proton number) configurations at the expense of a larger surface
energy.   These shell states may be related to the appearance of a central depression in the density of super heavy nuclei \cite{heavy} because of their large coulomb energies.
The appearance of shell states may significantly change transport properties such as the shear viscosity and shear modulus of neutron rich matter.

To determine the minimum free energy per baryon at a specified baryon density $n_B$, one must
search over the cell radius $R_c$.  When $R_c$ is large and
$n_B$ is beyond the neutron drip density $\sim 10^{11}$ g/cm$^3$ for neutron rich matter, one will need to take
into account a large number of levels, since the nucleon number is
$A = 4\pi R_c^3n_B/3$.  The fact that the matter is at finite
temperature will require one to consider even more levels.  For
example, when $n_B$ = 0.080 fm$^{-3}$, proton fraction $Y_p$ = 0.4, and $R_c$ =
23.5 fm, there are 4892 nucleons inside one WS cell and one needs
to include 427 (419) neutron (proton) levels. It is hard to
achieve self-consistency for the mean fields with a large number
of levels.  To ensure the convergence of the self-consistent
iterations for the mean fields, it is important to have a good
initial guess for the mean field potentials.  Here our strategy is
to employ the convergent potentials for a nucleus with a smaller
$R_c$ as the starting guess for a nucleus with a slightly larger
$R_c$. In this way, new shell states with large cell radii are
found.  These can minimize the free energy over a significant
density range.


\begin{figure}
 \centering
 \includegraphics[height=10cm,angle=-90]{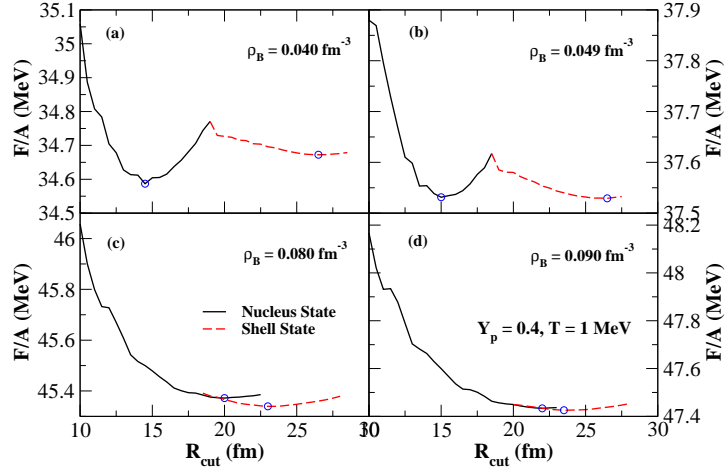}
 \caption{(Color online) Free energy per baryon versus Wigner-Seitz cell
 radius $R_c$ for baryon densities of $n_B$ = 0.040 (a),
0.049 (b), 0.080 (c), and 0.090 fm$^{-3}$ (d). The proton fraction
is $Y_p = 0.4$ and the temperature is $T$ = 1 MeV. The two circles
denote minima for the nucleus (small $R_c$) and shell states
(larger $R_c$).}
 \label{Yp0.4_mini}
\end{figure}

\begin{figure}
 \centering
 \includegraphics[height=10cm,angle=-90]{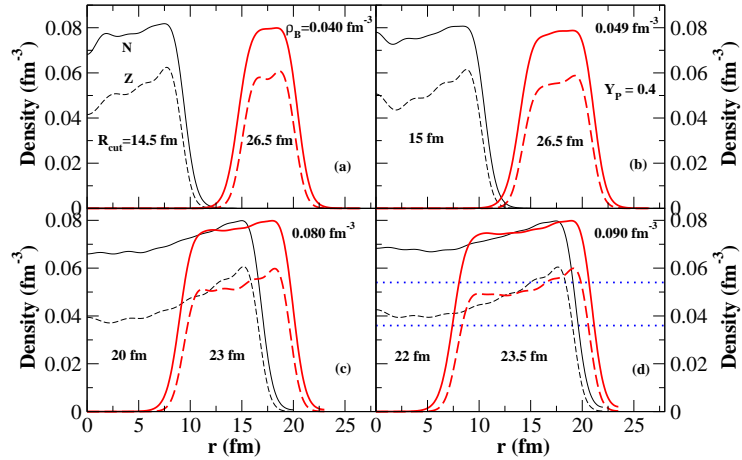}
 \caption{(Color online) The neutron (solid line) and
proton (dash line) density distributions for the Wigner-Seitz
cells with baryon densities of $n_B$ = 0.040 (a), 0.049 (b),
0.080 (c), and 0.090 fm$^{-3}$ (d). The proton fraction is $Y_p =
0.4$ and $T$ = 1 MeV. The red curves give density distributions
for shell states. The black curves give density distributions for
nucleus states. The corresponding cell radii are shown in each
panel. The dotted lines in the lower right panel are neutron
(upper) and proton (lower) densities for uniform matter.}
 \label{Yp0.4_density}
\end{figure}

We show in Fig.~\ref{Yp0.4_mini} an example of how shell states were found.  Here the free energy per baryon versus WS cell radius is displayed for baryon densities of 0.040, 0.049, 0.080, and 0.090 fm$^{-3}$ when $Y_p$ =0.4 and temperature $T$ = 1 MeV.  For each baryon density, there are two minima of free energy per baryon at different WS radii, denoted as open circles. The
neutron and proton density distributions of the WS cells with these
two radii are shown in Fig.~\ref{Yp0.4_density}. The first minimum
at smaller WS cell radius corresponds to a nucleus with a normal
density distribution. The second minimum at larger WS cell radius
corresponds to a shell shaped density distribution.  Here the nucleus has both outside and inside surfaces and the neutron and proton densities are only non-vanishing at intermediate $r$.  We call the first minimum a nucleus state and the second minimum a shell
state.  For the shell state, the sum of neutron and proton
densities (at intermediate radius $r$) is close to the saturation
density of nuclear matter $\sim$ 0.15 fm$^{-3}$, as one expects from nuclear saturation.

At $n_B$ = 0.049 fm$^{-3}$, the two minima of free energy per
baryon become degenerate, as shown in upper right panel of
Fig.~\ref{Yp0.4_mini}.  Below 0.049 fm$^{-3}$, the first minimum
with smaller cell radius corresponding to a normal nucleus is the
absolute minimum and therefore the equilibrium state ({\it e.g.},
panel a of Fig.~\ref{Yp0.4_mini}). On the other hand, above 0.049
fm$^{-3}$, the second minimum corresponding to a shell state is
the true equilibrium state ({\it e.g.}, panels c and d of
Fig.~\ref{Yp0.4_mini}). Therefore 0.049 fm$^{-3}$ is the critical
baryon density when the density distribution inside the WS cell
changes from a normal nucleus to a shell state.  In the lower
right panel of Fig.~\ref{Yp0.4_density}, we also show the uniform
neutron and proton density distributions by dotted lines when
$n_B$ = 0.090 fm$^{-3}$.  0.090
fm$^{-3}$ is in fact the transition density from a shell state to uniform
matter.


\begin{figure}
 \centering
 \includegraphics[height=10cm,angle=-90]{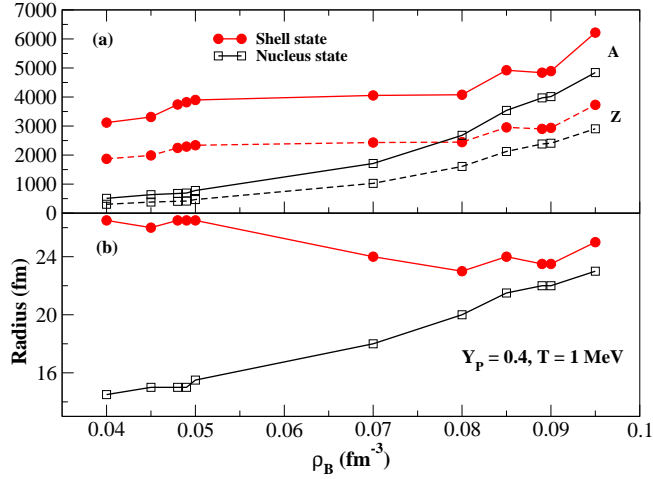}
 \caption{(Color online) The atomic and proton numbers (panel a),
 and the WS cell radii for nucleus (squares) and shell states (filled circles)
(panel b) shown as functions of the baryon density for $Y_p$ =
0.4.} \label{Yp0.4number}
\end{figure}

In Fig.~\ref{Yp0.4number}, the corresponding atomic and proton numbers 
$A$ and $Z$ for nucleus and shell states (upper panel), and the corresponding
WS cell radii (lower panel) are shown as functions of the baryon
density. At low baryon
densities, the WS cell radius of the shell state is about twice
that of the nucleus state, {\it e.g.}, 26 fm vs. 14 fm at $n_B$
= 0.040 fm$^{-3}$. The cell radii of shell states decrease with
increasing density, while those of nucleus states increase with
density.  They approach one another at high baryon densities: 24
fm vs. 22 fm at 0.090 fm$^{-3}$.  For both nucleus and shell
states, $A$ and $Z$ increase with increasing baryon density.

In this section our discussion is focused on $Y_p$ = 0.4 and $T$ = 1 MeV due to limit of space. 
The calculation has been extended to a large range of proton fractions and temperatures. We calculate the resulting
EoS for the inner crust (and core) of PNS at over 107,000 grid points in the proton fraction range $Y_P$ = 0 to 0.56. For the temperature
range $T$ = 0.16 to 15.8 MeV we cover the density range $n_B$ = 10$^{−4}$ to 1.6 fm$^{−3}$; and for the higher
temperature range $T$ = 15.8 to 80 MeV we cover the larger density range $n_B$ = 10$^{−8}$ to 1.6 fm${−3}$.

\section{Composition around neutrino-sphere}

Simulations of the core collapse supernovae and evolution of PNS depend heavily on the EoS at subnuclear density. Detailed information on the distributions of nuclei in the EoS table is important for neutrino-matter dynamics \cite{Burrows04}. Neutrinos radiate 99\% of the energy released in supernovae. Besides gravitational wave signals, neutrinos are the only messenger through which one can directly probe the EoS inside supernovae.  The neutrinosphere is the surface of last scattering before  $\nu$ or $\bar{\nu}$ escape.  The neutrinosphere is expected to be at a density around\ $10^{11}$ g/cm$^3$ and this is consistent with the available information from a handful
events in SN1987a  \cite{SN1987a,SN1987a2}.  The composition of matter at subnuclear density constrains the position of the neutrinosphere and influences the spectra of emitted neutrinos and antineutrinos. For example, in a recent study \cite{light07}, light nuclei with mass 2, 3 and 4 were found to have an important influence on the spectra of electron anti-neutrinos.


Nuclear statistical equilibrium (NSE) models treat
low density nuclear matter as a system of noninteracting nuclei in
statistical equilibrium, taking into account the binding
properties of heavy nuclei.  This has been widely used in nuclear
astrophysics \cite{NSE}. Recently, there have been several NSE based studies
of the supernova EoS, see for example Refs.
\cite{NSE09} and \cite{Hempel09}.

In this section, we discuss subnuclear density nuclear matter in a virial expansion for a nonideal gas, consisting of neutrons, protons, alpha particles, and 8980 species of heavy nuclei ($A\geq 12$) with masses from the finite-range droplet model (FRDM) \cite{FRDM}.  The virial results will cover  the density range $n_B$ = 10$^{-8}$ to 0.1 fm$^{-3}$, the temperature
range $T$ = 0.158 to 15.8 MeV, and the proton fraction range $Y_P$ = 0.05
to 0.56.  (For temperature higher than 15.8 MeV, matter is uniform and fully described in the RMF model.) The distribution of nuclei for given conditions is obtained in this approach, while the existing EoS tables of Lattimer-Swesty (L-S) \cite{LS} and H. Shen, Toki, Oyamatsu and Sumiyoshi (S-S) \cite{Shen98a,Shen98}, use a single heavy nucleus approximation.  The virial expansion is very similar to NSE when heavy nuclei are dominant, but includes important interactions between nucleons and light nuclei. The virial expansion is exact at low density limit when light elements dominate and fugacity of nucleons is small. Our virial EoS will be matched, using a thermodynamically consistent interpolation scheme, to the RMF EoS obtained in previous section to generate full EoS table for PNS and supernova.


We include the second order virial corrections among nucleons and alphas as in Ref.~\cite{HS05}, where the second order virial coefficients are obtained from scattering phase shift data.  Partition functions for heavy nuclei are included using the recipes of Fowler \etal \cite{Fowler78} (some calculations are presented using partition functions based on the recipe of Rauscher \etal \cite{Rauscher97}  for comparison in Ref. \cite{SHT10b} and the results are similar).  Equivalently, the partition functions can be considered as the sum of successive high orders of the virial expansion for heavy nuclei. There have been many studies on the level density and partition functions of hot nuclei in astrophysical environments.  For large scale
astrophysical applications, it is necessary to find both reliable
and computationally practical methods for the level density. Most of
these studies \cite{Gilbert65,Fowler78,Mazurek79,Engelbrecht91,Rauscher97} followed the original  non-interacting Fermi gas model of Hans Bethe \cite{Bethe36}.
For astrophysical nuclear reactions with temperature below a few
times 10$^{10} K$ \cite{Rauscher97,Rauscher00}, this
phenomenological approach gave excellent agreement with more
sophisticated Monte Carlo shell model calculations \cite{mc}, as
well as combinatorial approaches \cite{Engelbrecht91,Paar97}.  This justifies the application of the Fermi-gas description at and above the neutron separation energy.  For temperatures higher than a few times 10$^{10} K$, there are big ambiguities in
the values of the partition functions.  However, as suggested by some
authors \cite{NSE09} and supported by our own calculations,
these uncertainties have only a small impact on the thermodynamics of dense matter.

The effects of Coulomb interactions in the plasma can be estimated
by the plasma parameter $\Gamma_p = (Ze)^2/akT$, where $Z$
is the atomic number of the nucleus, $T$ is the temperature and $a$ is the
spacing between nuclei.  For matter at low density, $\Gamma_p$ is
smaller than one and the effect of Coulomb corrections is
small.  However for matter at higher density and when the dominant
species carry large charges, $\Gamma_p$ can be much greater than one and
the effect of Coulomb interactions should be taken into account. The Coulomb correction to the plasma has been studied analytically up to high $\Gamma_p$ by the cluster expansion \cite{coulomb}. Generally the correction due to
electron-ion interactions will reduce the free energy of the plasma and
eventually crystalize the matter at high density.  For simplicity, in this section
the Coulomb interactions between nuclei and electrons are included via a Wigner-Seitz approximation with effective ion spheres for each species of nuclei, wherein local electrical neutrality is maintained. This WS approximation for the Coulomb correction will be compared with a more rigorous Cluster expansion method.

\subsection{Average $A$ and $Z$}

\begin{figure}
 \centering
 \includegraphics[height=6.5cm,angle=-90]{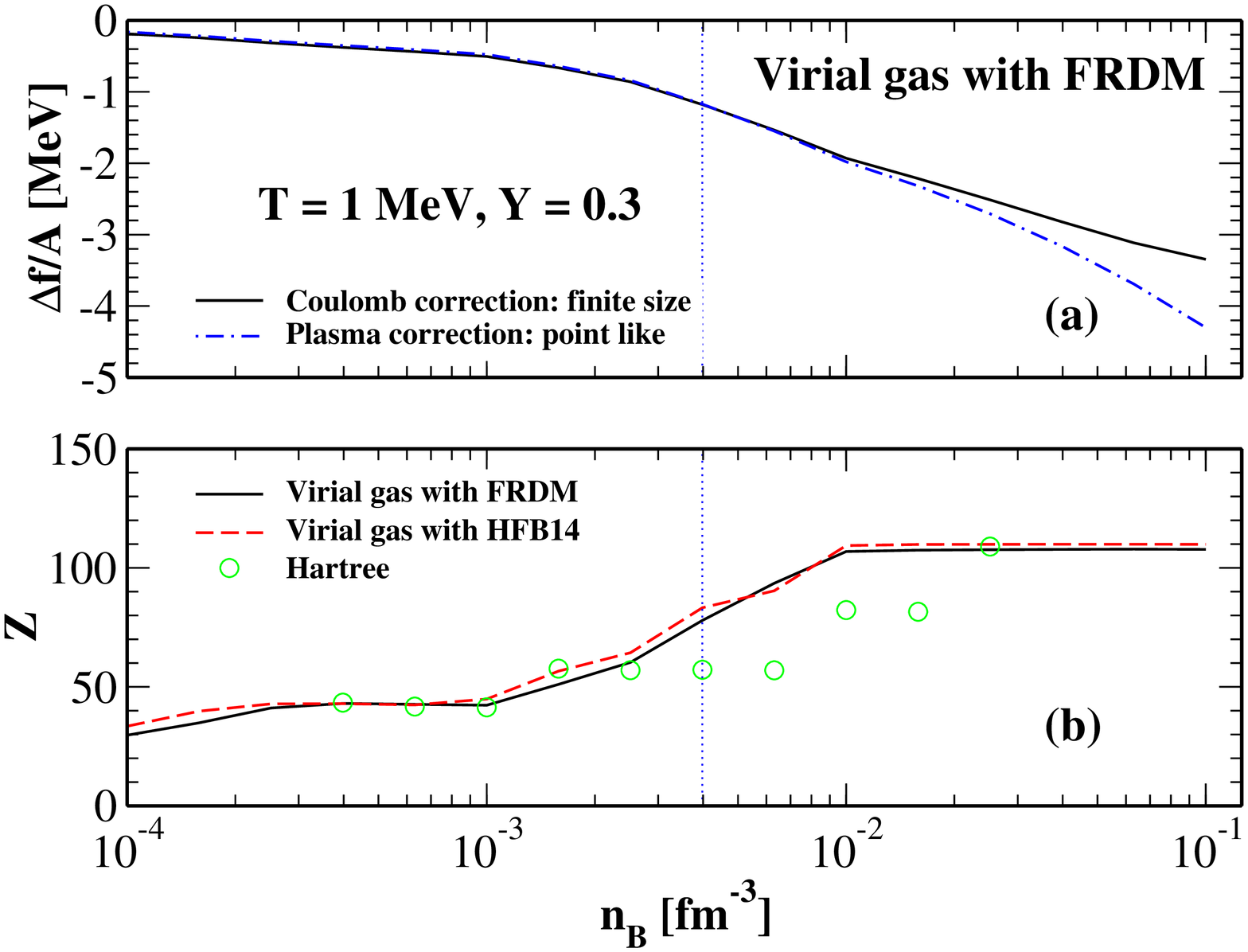}
 \includegraphics[height=6.5cm,angle=-90]{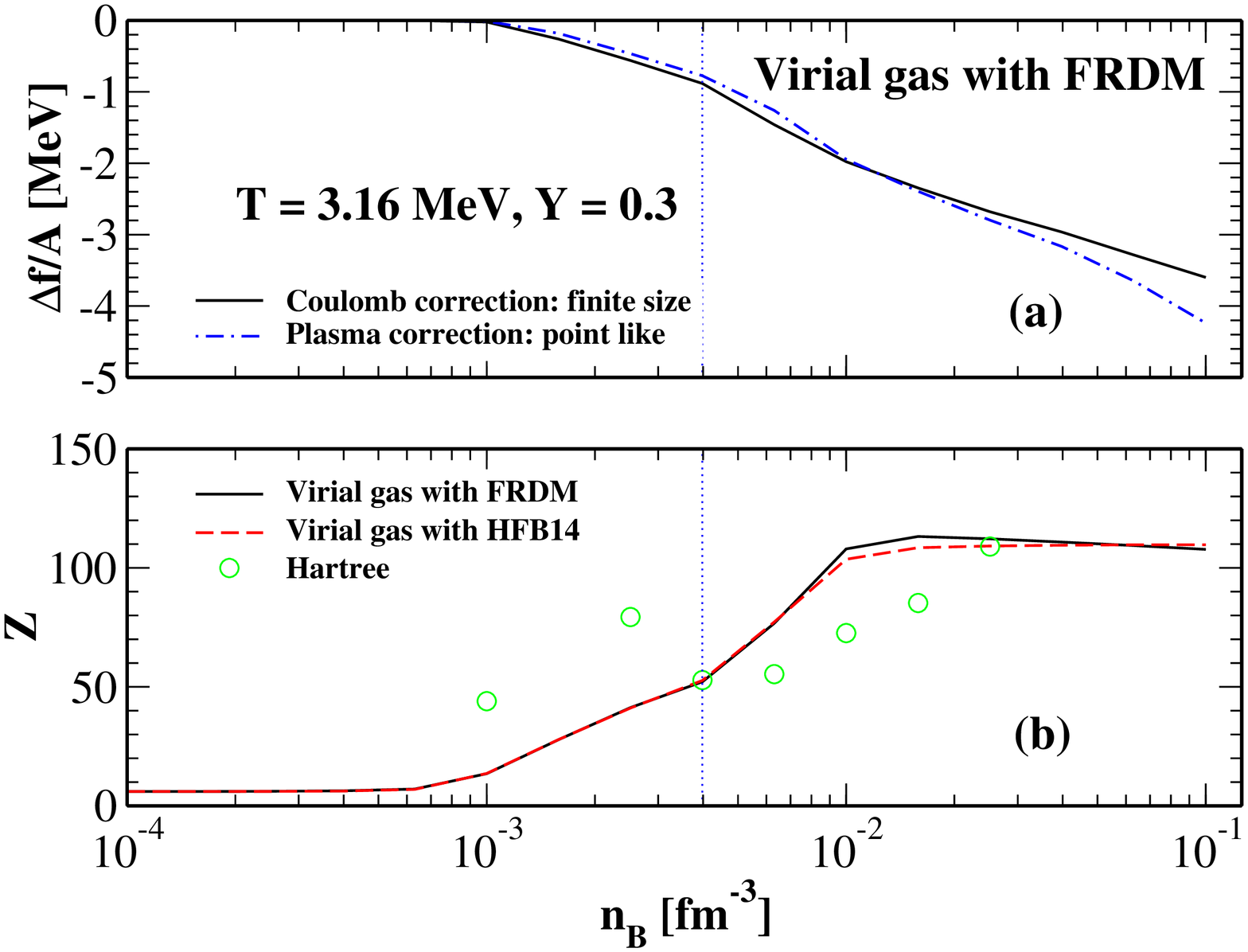}

\caption{(color on line) Upper: Coulomb corrections; lower: average charge number of heavy nuclei, in nuclear matter at $T$ = 1 (left panel) and 3.16 (right panel) MeV with $Y_P$ = 0.3.}\label{fig:cou_masstable}
\end{figure}

The lower panels in Fig. \ref{fig:cou_masstable} give the corresponding average charge number $Z$ from the virial EoS with either FRDM or HFB14 mass tables \cite{HFB14}, and from Hartree mean field results. Here temperature $T$ = 1 (left panel) and 3.16 (right panel) MeV with proton fraction $Y_P$ = 0.3.  The virial EoS with the two mass tables predict very similar values for $Z$.  Moreover, the virial EoS with either mass table gives very similar $Z$ to that from the Hartree mean field results at the transition density 
3.98$\times$10$^{-3}$ fm$^{-3}$ (blue dotted line). One can also identify several plateaus in the average $Z$, which correspond to closed shells (magic numbers). The fluctuation of $Z$ in Hartree results below the transition density (at $T$ = 3.16 MeV) is probably due to finite step error in search of cell size. We also studied free energy and transition density to Hartree mean field results, and found little dependence on the choice of mass tables.

In the two upper panels in Fig. \ref{fig:cou_masstable}, we compare the Coulomb energy correction in our virial expansion, with an analytic cluster expansion for the one-component plasma, Eq. (22) in \cite{coulomb}. The overall agreement for densities below the virial-Hartree transition is good, though at higher temperature the differences become larger. The reason is probably that we calculate the multi-component contribution to the Coulomb correction in the virial gas while we only used the average charge number in the analytic formula for the one-component plasma.

\subsection{Mass fractions of species}

The virial expansion gives the distribution of heavy nuclei, where 8980
species of nuclei are in thermal and chemical equilibrium with
free neutrons, free protons and alpha particles. This is an improvement over the L-S EoS and S-S EoS that both use a single-nucleus representation.

\begin{figure}
 \centering
 \includegraphics[height=8cm]{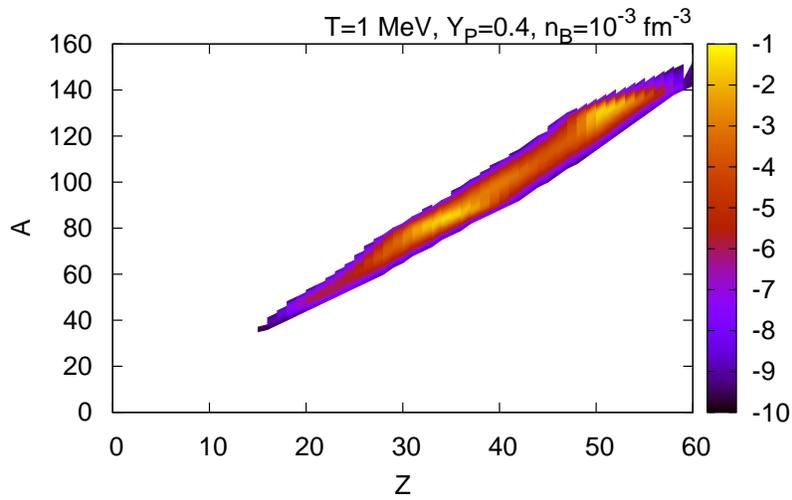}

\caption{Mass fraction of nuclei in the nuclear chart for matter
at $T$ = 1 MeV, $n_B$ = $10^{-3}$
fm$^{-3}$, and $Y_P$ = 0.4. Different colors indicate mass fraction in Log$_{10}$ scale. }\label{fig:dist_T1Y4n-3}
\end{figure}

\begin{figure}
 \centering
 \includegraphics[height=11cm,angle=-90]{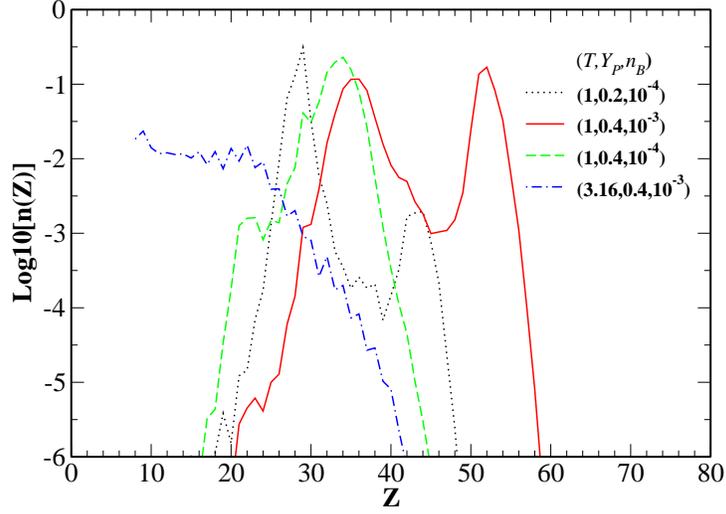}

\caption{Mass fractions of nuclei $n(Z)$ for nuclei with same proton number $Z$. The triplet in the parenthesis stands for ($T$/[MeV], $Y_P$, $n_B$/[fm$^{-3}$]).}\label{fig:z(n)}
\end{figure}

In Fig.~\ref{fig:dist_T1Y4n-3} we show mass fractions of different nuclei for matter with $T$ = 1 MeV, $Y_P$ = 0.4, and $n_B$ = $10^{-3}$ fm$^{-3}$.  The total mass fraction of heavy nuclei is close to unity and the distribution is centered around $Z$ = 35 and 50. The mass distribution of heavy nuclei in this high $Y_P$ case is a double-peaked Gaussian distribution, as shown in Fig.~\ref{fig:z(n)}, where $n(Z)$ is sum of the abundances of heavy nuclei with same proton number $Z$.

\begin{figure}
 \centering
 \includegraphics[height=6.5cm,angle=-90]{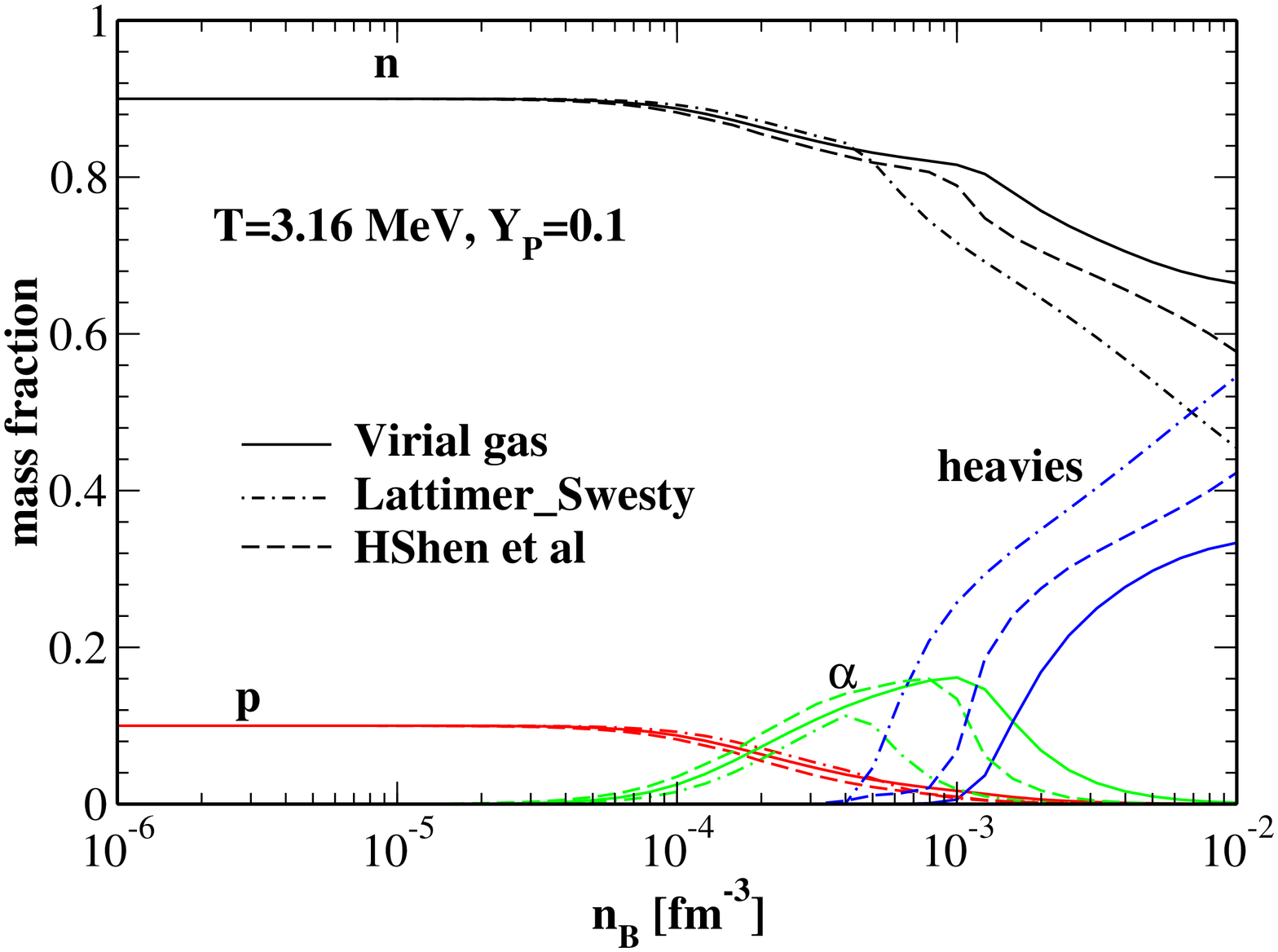}
 \includegraphics[height=6.5cm,angle=-90]{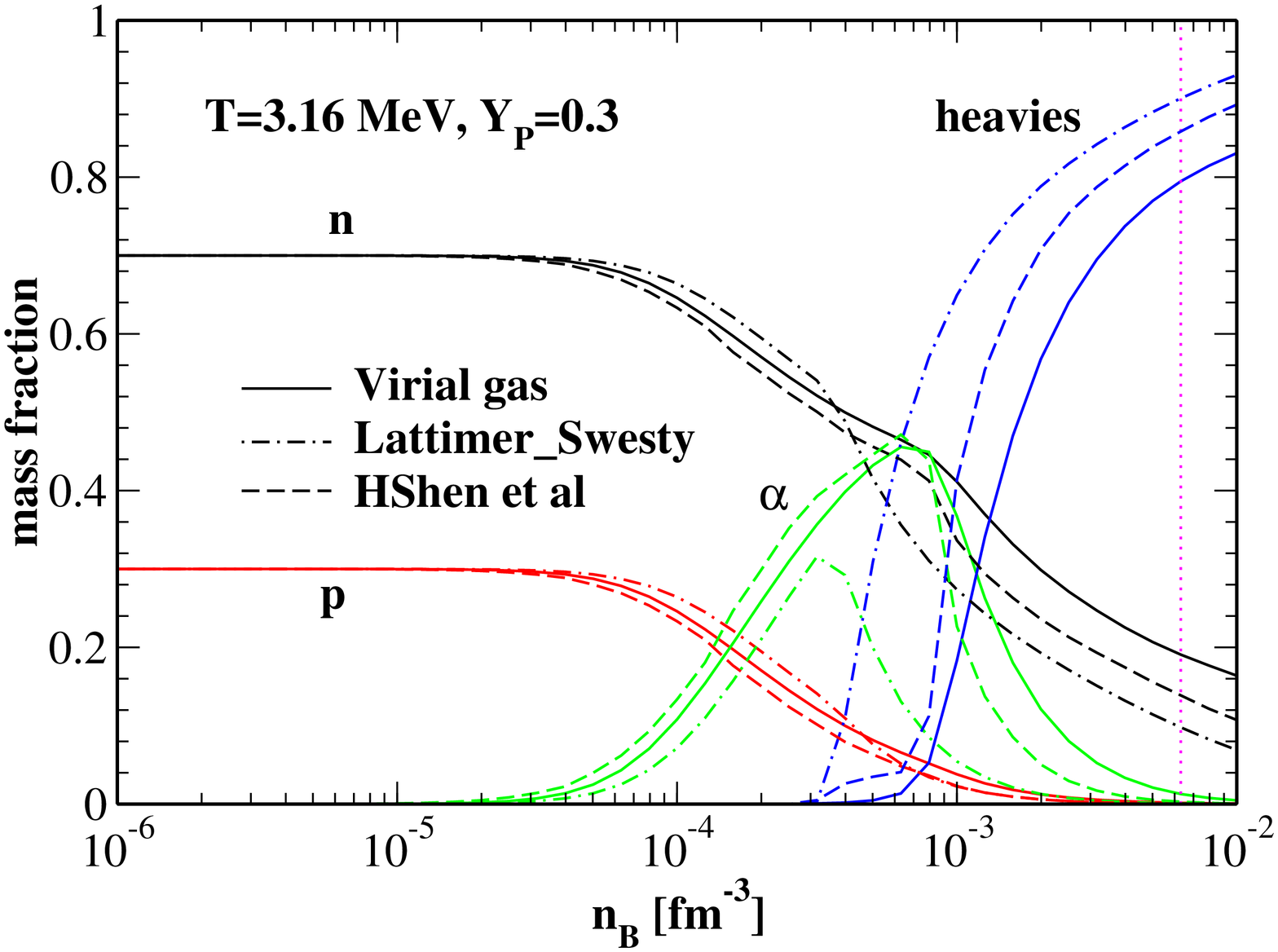}

\caption{Mass fractions of matter
at $T$ = 3.16 (right) MeV, $Y_P$ = 0.1 (left) or 0.3 (right) in our EoS (solid), Lattimer-Swesty EoS (dot-dashed) a nd H Shen \etal EoS (dashed).}\label{fig:compareabun}
\end{figure}

It is instructive to compare the composition of matter in virial EoS with the existing EoS tables, the L-S EoS and S-S EoS. The location of neutrinosphere in supernova is sensitive to the composition of matter and is important for the emitted neutrino spectra. Studies on collective flavor oscillation of neutrinos during their streaming outside neutrinosphere have already indicated sensitivity of neutrino flavor flip to the emitted neutrino spectra in neutrinosphere \cite{Duan10}. Below we will compare some examples for the composition of matter around neutrinosphere from virial EoS, the L-S EoS, and S-S EoS.

Fig.~\ref{fig:compareabun} shows the mass fractions of neutrons, protons, alpha particles and nuclei in matter at densities from 10$^{-6}$ to 10$^{-2}$ fm$^{-3}$. The matter has a temperature of 3.16 MeV and a proton fraction of 0.1 (left) or 0.3 (right), respectively. In the left panel for proton fraction of 0.1, free neutrons and protons dominate untill the density reaches 10$^{-4}$ fm$^{-3}$ in all three EoSs. Above 10$^{-4}$ fm$^{-3}$, alpha particles appear. The S-S EoS is close to our virial results at densities roughly below 10$^{-3}$ fm$^{-3}$.   The L-S EoS significantly underestimate X$_\alpha$ and this may be due to an error in the alpha particle binding energy.  Alpha particles have larger abundance and exists up to higher densities in our virial EoS than the other EoSs.  This is partly because the attractive interactions between neutrons and alpha particles in the virial expansion favors more alpha particle \cite{HS05}.  Heavy nuclei begin to appear around 4$\cdot$10$^{-4}$ fm$^{-3}$ in the L-S EoS, and at higher densities in the S-S EoS and our virial EoS.  Moreover, the L-S EoS predicts the largest abundance for heavy nuclei, while ours predicts the smallest abundance.  Free neutrons have the largest abundance in our virial EoS.  This is due to the strong attractive interaction between neutrons in the virial expansion which lowers the energy and enhances the abundance of neutrons.  Note in this $Y_P$ = 0.1 case the virial-Hartree transition happens at 0.0158 fm$^{-3}$. The right pane of Fig.~\ref{fig:compareabun}, for a different proton fraction of 0.3, has similar characteristics.  However here, alpha particles and heavy nuclei have much larger abundances than for the $Y_P$ = 0.1 case, since a higher proton fraction favors formation of nuclei. In this $Y_P$ = 0.3 case, the transition density from virial gas to Hartree mean field calculations occurs at 6.3$\cdot$ 10$^{-3}$ fm$^{-3}$ as indicated by the dotted line in the figure.

\section{Complete EoS for PNS}

In section 2 we used a relativistic mean field model to self-consistently calculate non-uniform matter at intermediate density and uniform matter at high density.  In section 3 we used a Virial expansion for nonideal gas of nucleons and nuclei to obtain the EoS at low densities.  Altogether these two EoS models cover the large range of temperatures, densities, and proton fractions.
Discussion of matching the two results can be found in Ref. \cite{SHT11}. There are 73,840 data points
from the Virial calculation at low densities, 17,021 data points from the nonuniform 
Hartree calculation, and 90,478 data points from uniform matter
calculations. The overall calculations took
7,000 CPU days in Indiana University's supercomputer clusters.

\begin{figure}
 \centering
 \includegraphics[height=10cm,angle=-90]{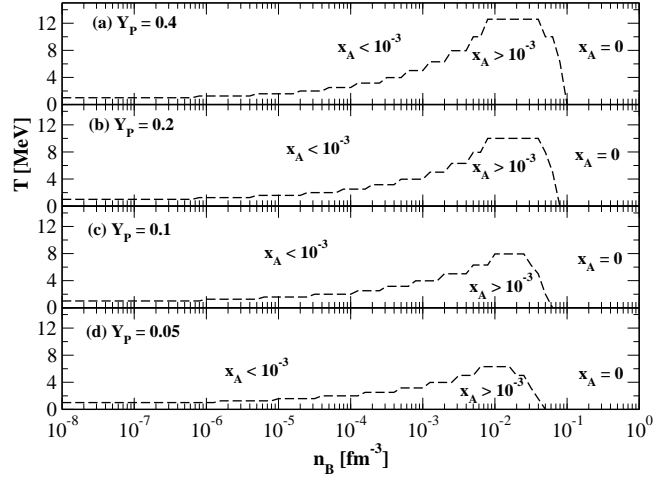}

\caption{Phase diagram of nuclear matter at different proton
fractions, 0.05, 0.1, 0.2 and 0.4. $X_A$ is mass fraction of heavy
nuclei with mass number $A$ $>$ 4.}\label{fig:phaseboundary}
\end{figure}

In Fig.~\ref{fig:phaseboundary}, we show the phase boundaries of
nuclear matter at different proton fractions $Y_P$ = 0.05, 0.1,
0.2 and 0.4. The mass fraction of heavy nuclei with mass number $A>4$ is $X_A$.  The boundary at low densities indicates when $X_A$ is greater or
less than 10$^{-3}$. The boundary at high densities indicates the
transition between non-uniform matter and uniform matter.  At very
low densities, the matter is dominated by free nucleons and alpha
particles.  As the density rises, heavy nuclei persist to higher temperatures.  Finally uniform matter takes over at sufficiently high density.  Figure~\ref{fig:phaseboundary} also shows that as the proton fraction rises, the temperature regime with
appreciable heavy nuclei grows and the transition density to
uniform matter increases.  The density for the nonuniform
matter to uniform matter transition has a weak temperature dependence.

We use a hybrid interpolation scheme to generate a full EoS table on a fine grid that is
thermodynamically consistent. The range of parameter spaces is shown in Table \ref{tab:phasespace2}.  This insures that the first law of thermodynamics is satisfied and that entropy is conserved during adiabatic compression.   Our EoS is an improvement over the existing  Lattimer-Swesty \cite{LS} and H. Shen \etal~ \cite{Shen98a,Shen98}, equations of state because our EoS includes thousands of heavy nuclei and is exact in the low density limit.

We also generated a second EoS based on the RMF effective interaction FSUGold \cite{Fsugold,SHO}, whereas our earlier EoS was based
on the RMF effective interaction NL3. The FSUGold interaction has a lower pressure at high
densities compared to the NL3 interaction. The original FSUGold interaction produces an EoS, that we call FSU1.7, that has a maximum neutron star mass of 1.7 solar masses. A modification in the high density EoS is introduced to increase the
maximum neutron star mass to 2.1 solar masses and results in a slightly different EoS that we call
FSU2.1. Finally, the EoS tables for NL3, FSU1.7 and FSU2.1 are available for download. 

\begin{table}
\centering \caption{Range of temperature $T$, density $n_B$, and proton
fraction $Y_p$ in the finely spaced interpolated EoS table.} \label{tab:phasespace2}\btab{cccc}
\hline \hline
Parameter & minimum & maximum & number of points \\
 \hline
T [MeV] & 0, 10$^{-0.8}$ & 10$^{1.875}$ & 109 \\

log$_{10}$(n$_B$) [fm$^{-3}$] &-8.0 & 0.175 & 328 \\

Y$_P$  & 0, 0.05  & 0.56 & 1(Y$_P$=0)+52 \\

\hline \etab
\end{table}

\section{Symmetry Energy in Various EoS}

The bulk properties of infinite nuclear matter have been collected in Table~\ref{Table1b}, for NL3, FSUGold, as well as a new effective RMF interaction IUFSU \cite{IUFSU}. One important distinction among various EoS is the slope of the symmetry energy at saturation density, $L/3\ =\ \rho_0 E'_{sym}(\rho_0)$.  The pressure around saturation density is proportional to $L$, which plays a crucial role both in the terrestial context where it affects the neutron density distribution in neutron rich nuclei and in astrophysics where it affects the structure and thermal evolution of neutron stars.

\begin{table}
\centering
\begin{tabular}{|l||c|c|c|c|c|}
 \hline
 Model & $\rho_{0}~({\rm fm}^{-3}) $ & $\varepsilon_{0}$~(MeV) 
           & $K_{0}$~(MeV) & $E_{sym}$~(MeV) & $L$~(MeV) \\
\hline
 \hline
 NL3    &  0.148  & $-$16.24 & 271.5 & 37.29 & 118.2 \\ 
 FSU    &  0.148  & $-$16.30 & 230.0 & 32.59 & 60.5  \\
IU-FSU & 0.155  & $-$16.40 & 231.2 & 31.30 & 47.2  \\

\hline
\end{tabular}
\caption{Bulk parameters characterizing the behavior of
              infinite nuclear matter at saturation density 
              $\rho_{_{0}}$. The quantities $\varepsilon_{_{0}}$ 
              and $K_{0}$ represent the binding energy per 
              nucleon and incompressibility coefficient of 
              symmetric nuclear matter, whereas $E_{sym}$ and 
              $L$ represent the energy and slope of the 
              symmetry energy at saturation density.}
\label{Table1b}
\end{table}

\begin{figure}
\centering
 \includegraphics[height=6.5cm,angle=-90]{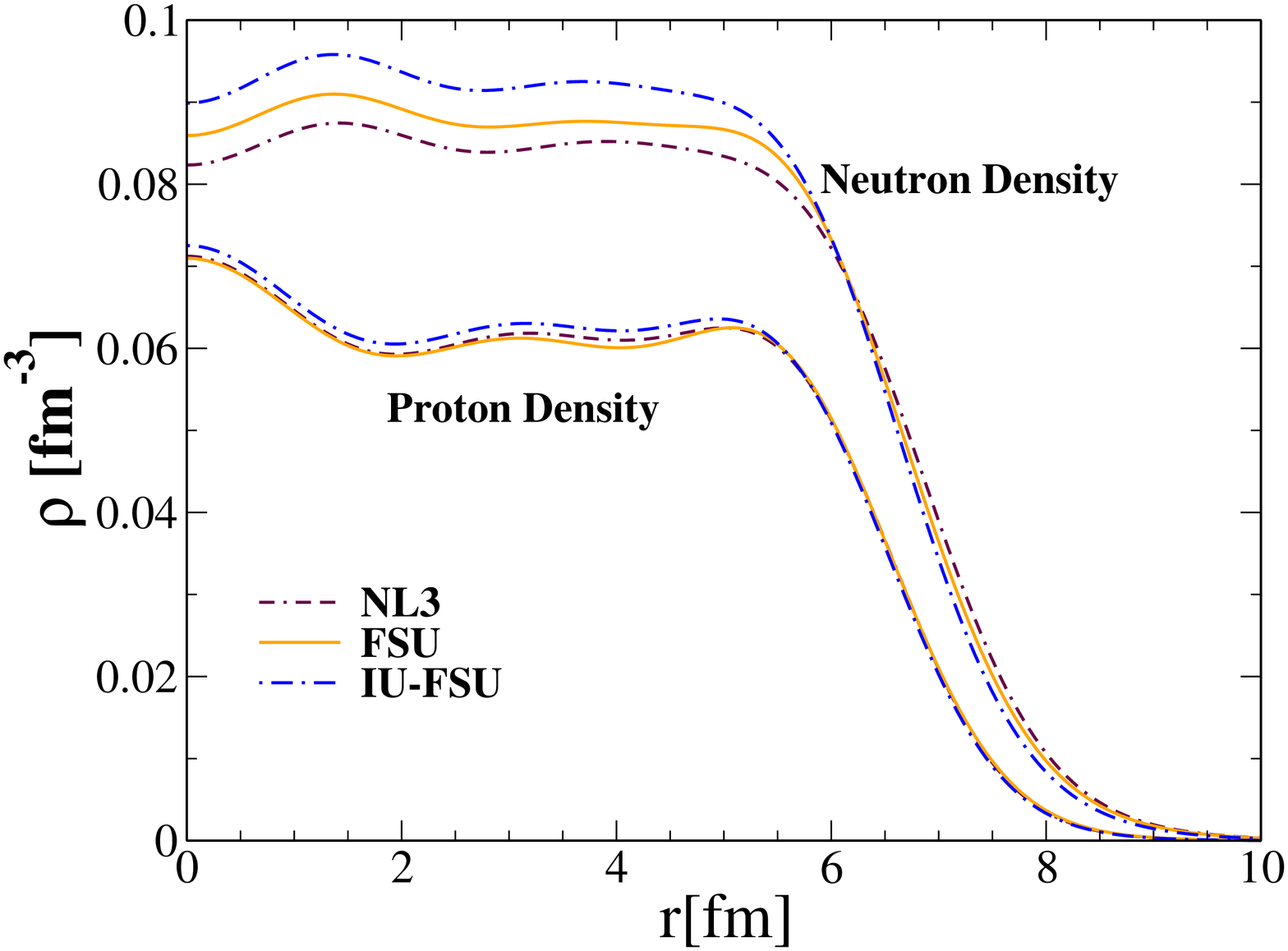}
\vspace{0.05in}
  \includegraphics[height=6.5cm,angle=-90]{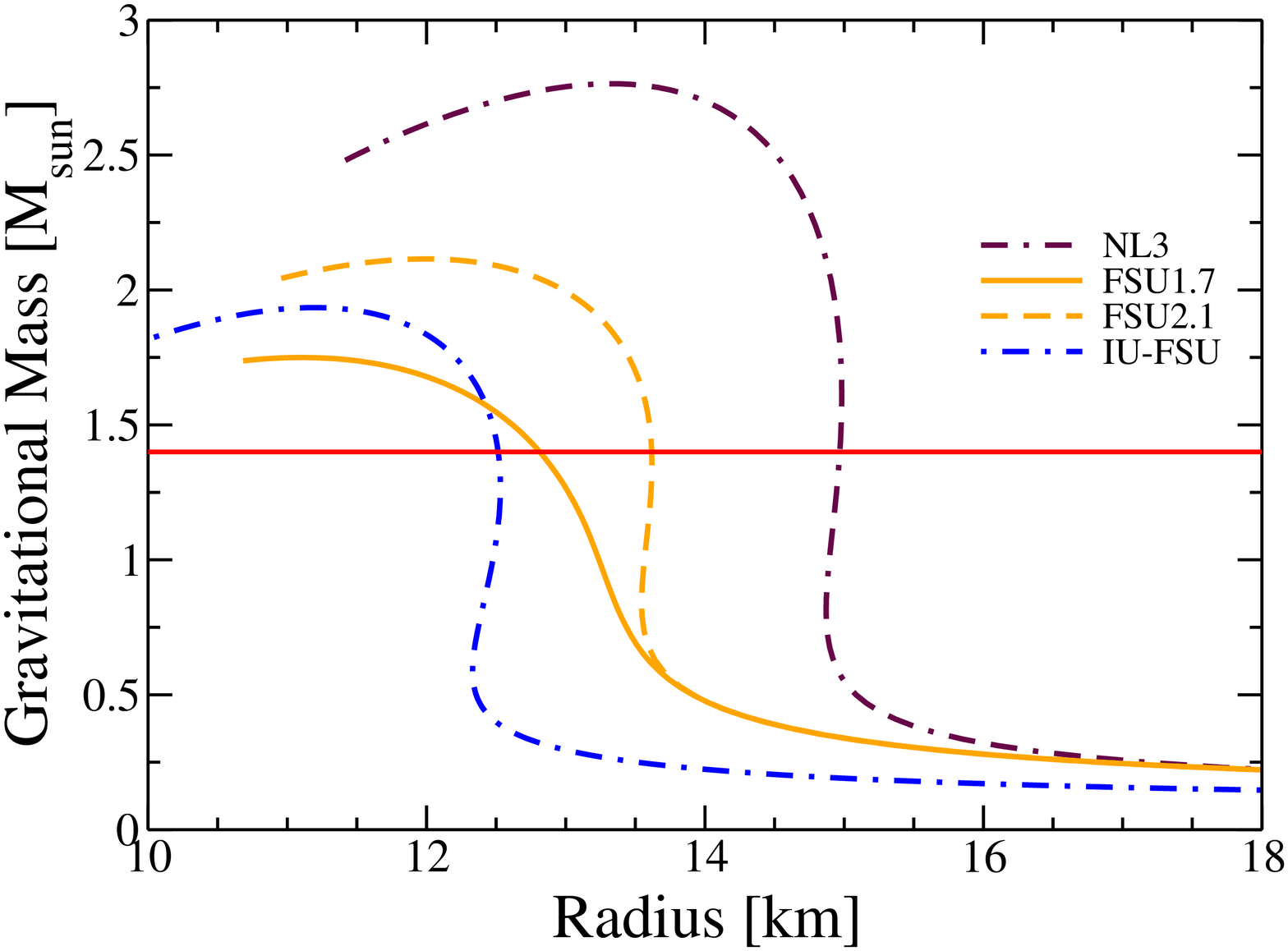}
\caption{(Color online) Left: model predictions for the proton and neutron densities of 
                ${}^{208}$Pb. Right: Neutron Star {\sl Mass-vs-Radius} relation predicted by 
 the relativistic mean-field models discussed in the text.}
\label{Fig1}
\end{figure}

Brown first realized the correlation between $L$ and the neutron skin thickness of $^{208}$Pb \cite{Brown00}, which has 126 neutrons and 82 protons. A larger pressure - due to larger $L$ inside the nucleus will push neutrons to the surface, therefore leads to a bigger neutron skin thickness. This is clearly demonstrated in left panel of Fig. \ref{Fig1}, where the proton and neutron densities inside  $^{208}$Pb are shown from several model predictions. The proton density is well constrained to 1\%. In contrast the neutron density has sizable variations among different model predictions. The values of $L$ are 118.2, 60.5, and 47.2 MeV for NL3, FSU, and IUFSU, respectively. The resulting neutron skin thickness is 0.28, 0.21, and 0.16 fm for NL3, FSU, and IUFSU, respectively. The pressure of neutron matter around saturation density also influences the radius of cold neutron star \cite{HP01,LP01}.  In the right panel of Fig. \ref{Fig1}, we show the neutron star mass-radius relation for various RMF models. For 1.4 solar mass neutron star, the corresponding radii are 15, 12.8, and 12.5 km for NL3, FSU, and IUFSU, respectively. FUS2.1 has a larger pressure than FSU1.7 at density above 0.2 fm$^{-3}$, and gives rise to larger radius for 1.4 solar mass neutron star, 13.6 km. Due to their common relation to the derivative of symmetry energy $L$, there exists a correlation between the neutron skin thickness and neutron star radius \cite{HP01}.

It's our goal to calculate EoS tables with different pressures.  This will allow one to correlate features of astrophysical simulations with EoS properties.  In this work we discussed two new EoS based on different RMF effective interactions NL3 and FSUGold. In future we will present a third EoS based on IU-FSU like effective interaction which has a softer symmetry energy $L$.

\section{Conclusions}

Equation of state of nuclear matter at finite temperature and its dynamics is the key to understand the evolution of proto-neutron star. The pressure at high densities determines how big and large a neutron star nature could make. The composition of matter in PNS and its dynamical response with neutrinos is important for the evolution of PNS. The emergent neutrino spectra from neutrino-sphere is crucial for the neutron fraction in the neutrino-driven wind and possible r-process that could make heavy nuclei beyond iron. The EoS of nuclear matter has been the focus of heavy ion collision experiments and a future heavy ion accelerator FRIB.

We construct several new EoS of nuclear matter for a wide range of temperatures, densities, and proton fractions. We employ fully microscopic relativistic mean field calculation for matter at intermediate density and high density, and the virial expansion of a nonideal gas (with nucleons and 8981 kinds of nuclei) for matter at low density. The EoS was obtained at over 180,000 grid points in 3-dimensional parameter spaces (temperature, density, and proton fraction). We used hybrid interpolation scheme to generate the final table, as shown in Table \ref{tab:phasespace2}. The thermodynamic consistency in our table is checked via usual adiabatic compression test, where the oscillation in entropy is limited to 1\% using the finer table Ref. \cite{SHT11,SHO}.

For nonuniform matter we find new shell states which minimize the free
energy per baryon over a significant density range. Shell states
have inside and outside surfaces and they can minimize the Coulomb
energy of high $Z$ (large proton number) configurations at the expense of a larger surface
energy. The appearance of shell states may significantly change transport properties such as the shear viscosity and shear modulus of neutron rich matter. The virial expansion gave broad distributions of nuclei in the nuclear chart, that are absent in two commonly used EOS tables, the Lattimer-Swesty EOS and H. Shen \etal EOS. These distributions of nuclei may be important for neutrino-matter interactions in supernova and the position of neutrino-sphere in supernova. The virial gas reduces to nuclear statistical equilibrium at low density and is exact in low density limit. Our new table is an improvement over existing EOS tables. We aim to provide detailed composition information in future EoS tables.  This can be important for neutrino interactions.

Due to the rather large uncertainties in the EoS at high density, particularly the density dependence of symmetry energy, we generated two EoS tables based on a stiff EoS at high density, NL3, and a softer EoS at low density, FSU1.7, and the modified FSU2.1 which could support a 2.1 solar mass neutron star. In future we will generate a third EoS based on IU-FSU like interaction, which is soft around saturation density but stiff at higher densities. Altogether these EoS tables could cover the uncertainties in the properties of nuclear matter at high density, and astrophysical simulations with them could identify observational phenomena with dinstinct nuclear matter property.

The property of neutron rich matter has yet to be fully understood, and currently under extensive studies both in experiments, such as PREX to measure the neutron skin thickness of $^{208}$Pb, and in theory, such as chiral effective theory calculation to third order perturbative expansion \cite{chiral} and quantum Monte Carlo method with uncertain 3-body forces \cite{qmc}. There are also important efforts to determine the EoS based on statistical analysis on the observational data such as X-ray burst on neutron stars \cite{Ozel10,bayesian}.

Probably an equally important problem for the PNS is the dynamical response of nuclear matter. Nuclear pasta could dominate in the composition of inner crust and neutrinos could scatter coherently on them \cite{pasta2,pasta3}. Dymanical response of nuclear matter around neutrino-sphere where light nuclei abound is important for the emitted neutrino spectra \cite{light07}. All these neutrino interactions have yet to be included in the studies of PNS evolution.

The author acknowledges supports by a grant from the DOE under contract DE-AC52-06NA25396 and the DOE topical collaboration to study "Neutrinos and nucleosynthesis in hot and dense matter" during write-up of this work.

\label{lastpage-01}

\end{document}